\documentclass[useAMS,usenatbib]{mn2e} \usepackage{graphicx} \title[On
the Cooling Trend of SGR~0526$-$66]{On the Cooling Trend of SGR~0526$-$66}

\author[T. G\"uver, E. G\"o\u{g}\"u\c{s} and F. \"Ozel]
{Tolga G\"uver$^{1}$, Ersin G\"o\u{g}\"u\c{s}$^{1}$ and Feryal \"Ozel$^{2}$ \\
  $^{1}$Sabanc\i~University, Faculty of Engineering and Natural
  Sciences, Orhanl\i ~Tuzla 34956 Istanbul Turkey \\
  $^{2}$Department of Astronomy, University  of Arizona, 933 N. Cherry
  Ave., Tucson, AZ 85721}

\begin{document}
\date{}
\pagerange{\pageref{firstpage}--\pageref{lastpage}} \pubyear{2012}
\maketitle
\label{firstpage}
\newcommand{\sgr}{SGR~0526$-$66~}

\begin{abstract}

  We   present  a   systematic  analysis   of  all   archival  Chandra
  observations of the  soft-gamma repeater SGR~0526$-$66.  Our results
  show that the X-ray flux of  \sgr decayed by about 20\% between 2000
  and 2009.  We employ  physically motivated X-ray spectral models and
  determine the effective temperature and the strength of the magnetic
  field  at  the  surface  as  kT~=~0.354$_{-0.024}^{+0.031}$~keV  and
  B~=~(3.73$^{+0.16}_{-0.08}$)$\times$10$^{14}$~G,    respectively.
  We find  that the effective temperature remains  constant within the
  statistical uncertainties  and attribute the decrease  in the source
  flux to a  decrease in the emitting radius.   We also perform timing
  analysis to measure the evolution  of the spin period and the period
  derivative over the nine year  interval. We find a period derivative
  of   \.P~=~(4.0~$\pm$0.5)$\times$10$^{-11}$~s~s$^{-1}$,  which
  allows us to  infer the dipole magnetic field  strength and compare
  it with  the one determined spectroscopically.   Finally, we compare
  the effective  temperature of \sgr with the  expected cooling trends
  from magnetized neutron stars  and suggest an initial magnetic field
  strength of 10$^{15-16}$ G for the source.

\end{abstract}


\section{Introduction}

Soft Gamma Repeaters (SGRs) and Anomalous X-ray Pulsars (AXPs) are the
prime representatives  of magnetars --  a class of neutron  stars that
are thought to  be powered by the decay  of their superstrong magnetic
fields (B$\sim$10$^{14}$--10$^{15}$~G, Duncan \& Thompson 1992). Aside
from emitting  energetic bursts  in hard X-rays  and soft  gamma rays,
magnetars  are bright  X-ray  sources (L$_x$$\sim$10$^{33}$--10$^{36}$
erg~s$^{-1}$),   emitting  pulsed   X-rays   either  persistently   or
episodically  (see Woods  \&  Thompson 2006  and  Mereghetti 2008  for
detailed reviews).  X-ray  observations of most of SGRs  and AXPs over
the last  thirty years  have shown that  these sources are  not steady
emitters,  exhibiting flux variability  generally in  conjunction with
bursting activities (see e.g., Rea \& Esposito 2011).

All known  SGRs have  been discovered when  they went into  a bursting
phase,  during  which they  emit  repeated  energetic  bursts in  soft
gamma-rays. SGR~0526$-$66  was not an exception: it  was discovered as
the first SGR with an extremely bright gamma-ray burst on 1979 March 5
that  was followed  by  a tail  emission  clearly pulsating  at 8.1  s
(Mazets et  al.  1979).  The burst location  is within  the supernova
remnant, N49  in the Large Magellanic  Cloud (Evans et  al. 1980). The
source exhibited  16 more  bursts until 1983  April 5 with  much lower
intensity  than that of  the March  5th event  (Aptekar et  al. 2002).
Since then, no bursts have been detected from SGR~0526$-$66.

Due to its large distance and  the fact that the source is embedded in
a supernova remnant, spectral and timing studies of SGR~0526$-$66 have
been limited.  The persistent  X-ray source was identified with ROSAT,
yielding the earliest view of the  source as it is in burst quiescence
(Rothschild,  Kulkarni \& Lingenfelter  1994). SGR~0526$-$66  has been
the target of two Chandra X-ray Observatory (CXO) observations in 2000
and  2001.  Thanks  to the  superb  angular resolution  of CXO,  X-ray
spectroscopy and timing investigations of the persistent source became
possible. Based on these observations, spin period of $\approx$~8.04 s
was    measured     and    a    tentative     \.{P}    of    $\approx$
6.5$\times$10$^{-11}$~s~s$^{-1}$ was inferred  (Kulkarni et al. 2003).
X-ray spectrum of  the SGR~0526$-$66 could be described  well with the
empirical model of  an absorbed blackbody plus a  power law component.
The  resulting power law  index was  much steeper  than that  of other
SGRs, while it was similar to the power law indices of AXPs, which led
Kulkarni  et  al.   (2003)  to  suggest that  SGR~0526$-$66  is  in  a
transition from the SGR-like phase to an AXP-like phase.

SGR~0526$-$66 and  its associated supernova remnant,  N49 was recently
observed with XMM-Newton. Tiengo et al. (2009) reported that the X-ray
spectrum of the source does  not show any significant changes compared
to   XMM-Newton  observations   performed   in  2000   and  2001   and
SGR~0526$-$66  is still  a bright  persistent source,  emitting  at an
X-ray  luminosity  of $\approx$~4$\times$10$^{35}$~erg~s$^{-1}$.   Very
recently, Park  et al.   (2012) provided an  in depth analysis  of the
supernova remnant  using four Chandra observations  that were obtained
in    2009.    They    refined    the   Sedov    age    of   N49    as
$\tau_{\rm Sed}$~$\approx$~4800 yr.  Park et  al. (2012) also analysed the
X-ray spectra of \sgr using  phenomenological models, such as the sum
of a  blackbody and  a power-law or  two blackbody models.   They find
that  the  X-ray flux  of  \sgr  varied by  about  15  \% between  the
observations  performed  in  2000/2001  and 2009.   Further  including
archival ROSAT observations they detect a decay in X-ray flux by about
$\approx$~20-30\% for the last 17 years.

Here, we  report on  our systematic analysis  of all  archival Chandra
observations of  \sgr to unveil the X-ray  spectral characteristics of
the  source when it is  in deep  burst quiescence.  We model  the X-ray
spectra   with  the  Surface   Thermal  Emission   and  Magnetospheric
Scattering  (STEMS)  model (G\"uver  et  al.   2007,  2008, 2011),  to
determine  the  effective surface  temperature,  the surface  magnetic
field strength, and the long term X-ray flux of SGR~0526$-$66. We also
perform timing analysis to construct  the spin period evolution of the
source    and   obtain    the   inferred    dipole    magnetic   field
strength. Finally,  we compare our  results with the  expected cooling
trends from magnetized neutron stars.

\section{Observations and Data Analysis}

We list  the observations used in  our study in  Table \ref{obs}. Note
that   there  are   two   more  Chandra   observations  that   contain
SGR~0526$-$66 in the field of  view. However, the source was observed
off-axis by  about 4"  in one of  the observations (ObsID:  1041). The
effects  of  vignetting do  not  allow  a  reliable source  extraction
without  any  contribution  from  the  supernova  remnant.  The  other
observation (ObsID: 2515) has only 7 ks of effective exposure, that is
too   short    to   provide   high    enough   signal-to-noise   X-ray
spectrum.  Therefore,   we  exclude  these  two   pointings  from  our
investigations.

\begin{table}
\center
\caption{Chandra Observations of the SGR~0526$-$66.}
\begin{tabular}{ccc}
\hline
Date &  Observation & Exposure \\
     &  ID          & (ks) \\
\hline
2000-01-04  & 747   & 43.9 \\
2001-08-31 & 1957  & 53.4 \\ 
2009-07-18 & 10123 & 28.2 \\
2009-07-31 & 10808 & 30.2 \\
2009-09-16 & 10807 & 27.3 \\
2009-09-19 & 10806 & 27.9 \\
\hline
\end{tabular}
\label{obs}
\end{table}

\begin{figure*}
  \centering
\includegraphics[scale=0.75]{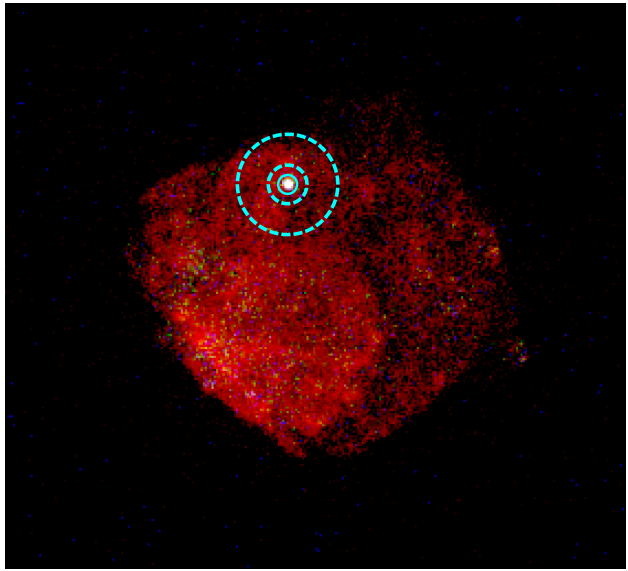}
\caption{True color  image of the  supernova remnant N49.   Red, green
  and  blue  corresponds to  the  0.2-2.0,  2.0-4.0  and 4.0-10.0  keV
  ranges,  respectively.  Source (solid  line) and  SNR (dashed
  lines) extraction regions are also shown in cyan.}
\label{image}
\end{figure*}

We reprocessed  all of the  observations with the CIAO  software suit,
version  4.2  and  CALDB  4.3.0  using the  recently  introduced  {\it
  chandra\_repro}\footnote{http://cxc.harvard.edu/ciao/ahelp/chandra\_repro.html}
tool. This tool automates the creation of new bad pixel file and level
2 event  file. We  applied barycentric correction  to each  event file
using the {\it axbary} tool. To extract source spectra and lightcurves
we selected a circular region centering the neutron star with a radius
of 2\arcsec,  as shown in  Figure \ref{image}.  For the  background we
used  an annular region  that is  centered on  the coordinates  of the
neutron star and  covers the range starting from  2.5 to 5\arcsec from
the center.  Finally,  we selected data from a  larger annular region:
centered on the neutron star and covering the range starting from 4 to
11\arcsec,  to generate an  X-ray spectrum  of the  SNR and  obtain an
independent measurement  of the  hydrogen column density,  as detailed
below.   Source  and  SNR  extraction  regions  are  shown  in  Figure
\ref{image}. We  extracted X-ray  spectra using the  {\it specextract}
tool  and created  response files  with the  {\it mkacisrmf}  and {\it
  mkarf} tools.  We  grouped each spectrum to have  at least 50 counts
per spectral channel.

\section{X-ray spectral analysis}
\label{spec}

We fit the data with  XSPEC version 12.5.1n, using the Surface Thermal
Emission  and  Magnetospheric   Scattering  and  assumed  a  reference
gravitational redshift  of 0.306 for the neutron  star.  We calculated
the unabsorbed  fluxes in the  0.5 to 6.5  keV energy range  using the
{\it cflux} model in XSPEC.  The  distance of the SGR is assumed to be
48.1 kpc  (Macri et al. 2006).  Uncertainties  reported throughout the
paper  correspond  to 68\%  confidence  limits  of parameters,  unless
indicated otherwise.

Uncertainties  in  the  amount  of interstellar  absorption  of  X-ray
photons  often hampers  efforts to  precisely determine  the intrinsic
spectral  shape of these  sources. Furthermore,  as a  free parameter,
hydrogen   column  density  becomes   strongly  correlated   with  the
determined temperature  of the X-ray source  essentially affecting the
resulting best fit  value (see, e.g.  Durant \&  van Kerkwijk 2006 and
G\"uver  et  al.  2012  for  a  detailed  discussion).  Therefore,  an
accurate  estimation of  the  hydrogen column  density  is crucial  to
deduce the spectral characteristics of  the source. The fact that \sgr
is embedded in the supernova remnant N49 presents an advantage in this
respect because it is possible to obtain an independent measurement of
the  hydrogen column  density using  the  SNR.  For  that purpose,  we
extracted  a  SNR  spectrum  from  the longest  exposure  and  modeled
it. Similar to the earlier studies of N49 (Park et al.  2003; Bilikova
2007; Tiengo et al.  2009; Park et al.  2012), we fit the remnant with
a two component model, consisting of two plane-parallel shocked plasma
functions  (Borkowski  et al.   2001,  {\it  vpshock}  in XSPEC)  both
absorbed  by the interstellar  medium, assuming  a solar  abundance as
given by Anders \& Grevesse (1989). We obtained an adequate fit with a
$\chi^{2}_{\nu}$ = 1.44  for 99 degrees of freedom.   We note that the
relatively high $\chi^{2}$ is mostly  due a small number of individual
energy bins, which could be emission lines that can not be resolved by
the ACIS-S detector.  Our best  fit parameters are: $N_{\rm H} = (0.15
\pm  0.03)\times  10^{22}$~cm$^{-2}$,  $kT_{1}  =  0.54~\pm~0.01$~keV,
$\tau_{1}   >    3.6   \times   10^{13}$~s~cm$^{-3}$,   $kT_{2}~=~1.11
\pm0.4$~keV,  $\tau_{2}~=~(1.21 \pm 1.1)  \times 10^{10}$~s~cm$^{-3}$,
where  $kT_{1}$,  $kT_{2}$,  $\tau_{1}$  and  $\tau_{2}$  denotes  the
temperature  of the  plasma  and  the upper  limit  on the  ionziation
timescale for  each component.  The elemental abundances  we found are
as               follows:              Ne/Ne$_{\sun}~=~0.90~\pm~0.10$,
Mg/Mg$_{\sun}~=~0.64~\pm~0.10$,         Si/Si$_{\sun}~=~0.66~\pm~0.08$,
S/S$_{\sun}~=~1.01~\pm~0.25$,  and Fe/Fe$_{\sun}~=~0.38~\pm~0.04$, all
the  other  abundances  were  set  to solar  values.   These  results,
especially the hydrogen column  density, are in general good agreement
with the  results of Park et  al.  (2012). We use  the hydrogen column
density we inferred from this  fit as a fixed parameter when analyzing
the spectra of SGR~0526$-$66 in the remainder of the paper.

\begin{figure*}
\centering
\includegraphics[scale=0.35]{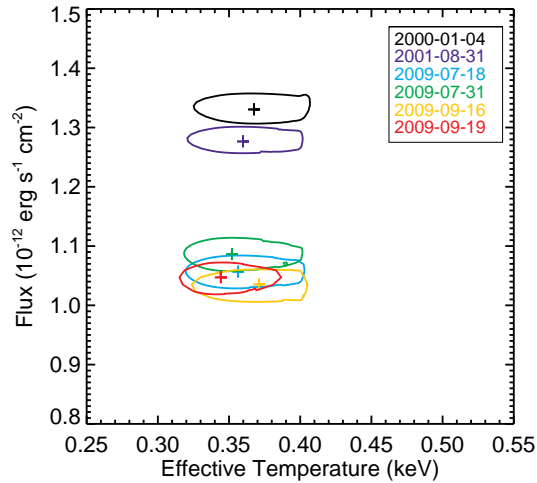}
\caption{68\% confidence  contours of unabsorbed 0.5 - 6.5
  keV  flux and  surface effective  temperature values  for individual
  data  sets when all other parameters were linked  between different
  observations.}
\label{kt_flux_conf}
\end{figure*}

We simultaneously  fit all  six spectra with  the STEMS  model.  STEMS
model  assumes  a  fully  ionized  and  strongly  magnetized  hydrogen
atmosphere on  the surface of  the neutron star, which  determines the
general spectral  characteristics of its X-ray  emission (\"Ozel 2001,
2003).   In  the  magnetosphere,  these surface  photons  are  further
scattered by  mildly relativistic charges (Lyutikov  \& Gavriil 2006).
The  model has  four parameters:  the effective  temperature, magnetic
field  strength at  the surface  as  well as  the resonant  scattering
optical depth  and the  velocity of charged  particles in  the neutron
star magnetosphere  (see e.g.,  G\"uver et al.   2007, 2008,  2011 for
details).   First, we  allowed all  STEMS  parameters to  vary in  all
observations.  Such  a fit resulted  in a $\chi^{2}/dof$ of  1.074 for
515 degrees of freedom (dof).  Resulting effective surface temperature
($\sim  0.35$~keV),   magnetic  field  strength   ($\sim  3.61  \times
10^{14}$~G), the magnetospheric scattering optical depth ($\sim 5.66$)
and  the  average  particle   velocity  ($\sim  0.55$)  did  not  show
statistically significant  or systematic variations  between different
observations.  However,  due to  rather low signal  to noise  ratio of
individual  X-ray spectra,  the error  in individual  model parameters
were  large.  In  order  to  obtain more  constrained  results on  the
effective  temperature (therefore, the  apparent emitting  radius) and
follow its time evolution, we linked the model parameters for magnetic
field  strength,  magnetospheric  scattering  optical  depth  and  the
average particle velocity.  This  way, we obtained a $\chi^{2}$/dof of
1.086/530.  We note here that, despite adding more dof to the fit, the
change in  the fit statistics was negligible,  further indicating that
the other parameters do not show statistically significant variations.
We  obtained the  following best  fit parameter  values when  only the
surface effective temperature and  model normalization were allowed to
vary  between  observations:   the  surface  magnetic  field  strength
B$=(3.74^{+0.11}_{-0.14})\times10^{14}$~G,    magnetospheric   optical
depth to resonant  scattering $\tau=5.37^{+0.58}_{-0.44}$, the average
velocity of  particles $\beta=0.52 \pm  0.03$.  We then used  the {\it
  cflux} model to calculate the unabsorbed X-ray flux in the 0.5$-$6.5
keV range  for each observation.   We present these  flux measurements
together  with  the  effective  surface temperature  values  for  each
observation  in   Table  \ref{kt_flux}.    We  also  show   in  Figure
\ref{kt_flux_conf}  the  68\%  confidence  contours  for  the  surface
effective temperature and flux values.  Both Figure \ref{kt_flux_conf}
and  Table \ref{kt_flux}  show that  although the  X-ray flux  of \sgr
decreases  by a  factor of  20\% over  the last  9 years  the inferred
temperature values are fairly constant during this period.

\begin{table*}
\center
\caption{X-ray spectral fit results for SGR~0526$-$66.}
\begin{tabular}{ccccc}
\hline
Date &Unabsorbed Flux$^{a}$ & $kT_{\rm eff}$$^{a}$ &  Unabsorbed Flux$^{b}$ & Radius$^{b}$ \\
& (10$^{-12}$ erg s$^{-1}$ cm$^{-2}$) & (keV) & (10$^{-12}$ erg s$^{-1}$ cm$^{-2}$) & (km)\\
\hline
  2000-01-04 &  1.33$\pm$0.02 & 0.37$_{-0.04}^{+0.04}$ & 1.31$\pm0.02$& 13.33$\pm$1.56\\
  2001-08-31 &  1.28$\pm$0.02 & 0.36$_{-0.04}^{+0.04}$ & 1.26$\pm0.01$& 13.09$\pm$1.54\\
  2009-07-18 &  1.06$\pm$0.02 & 0.36$_{-0.04}^{+0.05}$ & 1.04$\pm0.02$& 11.88$\pm$1.37\\
  2009-07-31 &  1.09$\pm$0.03 & 0.35$_{-0.04}^{+0.05}$ & 1.07$\pm0.02$& 12.05$\pm$1.39\\
  2009-09-16 &  1.04$\pm$0.02 & 0.37$_{-0.05}^{+0.03}$ & 1.00$\pm0.02$& 11.71$\pm$1.35\\
  2009-09-19 &  1.05$\pm$0.02 & 0.34$_{-0.03}^{+0.04}$ & 1.03$\pm0.02$& 11.85$\pm$1.37\\
\hline
\end{tabular}
\footnotesize{ \begin{flushleft} $^{a}$ Calculated from each individual data set when all the model parameters were linked between the observations, except for the effective temperature and normalizations. \\
    $^{b}$ Calculated from each individual data set when all the model parameters were linked between the observations, except for normalizations.
\end{flushleft}}
\label{kt_flux}
\end{table*}

\begin{figure*}
\centering
\includegraphics[scale=0.5, angle=270]{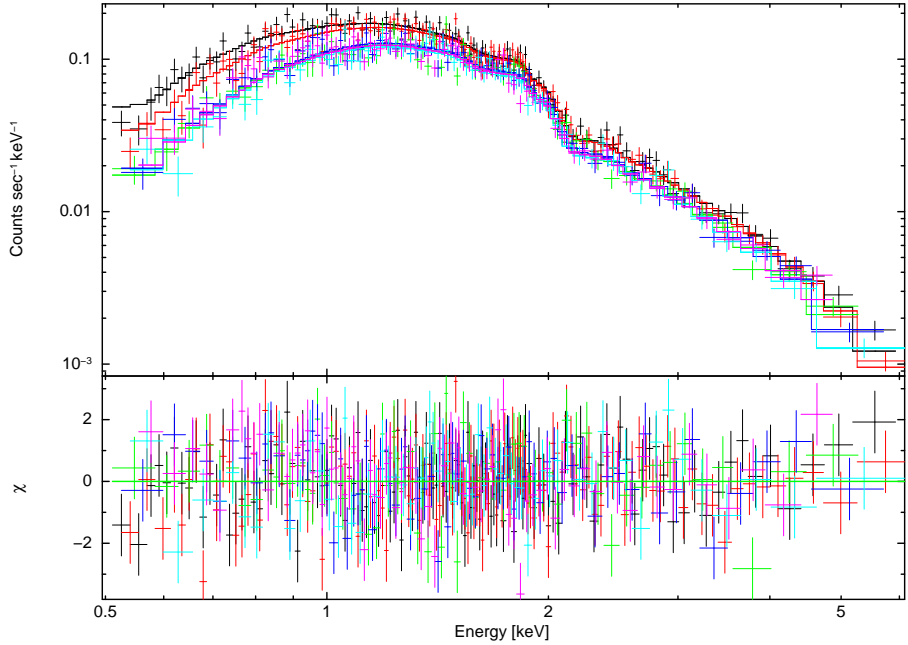}
\caption{All six  X-ray spectra  of SGR~0526$-$66 obtained  since 2000
  (crosses),  best fitting  model  curves (solid  lines)  and the  fit
  residuals (lower panel).}
\label{best_fit}
\end{figure*}

\begin{figure*}
\centering
\includegraphics[scale=0.3]{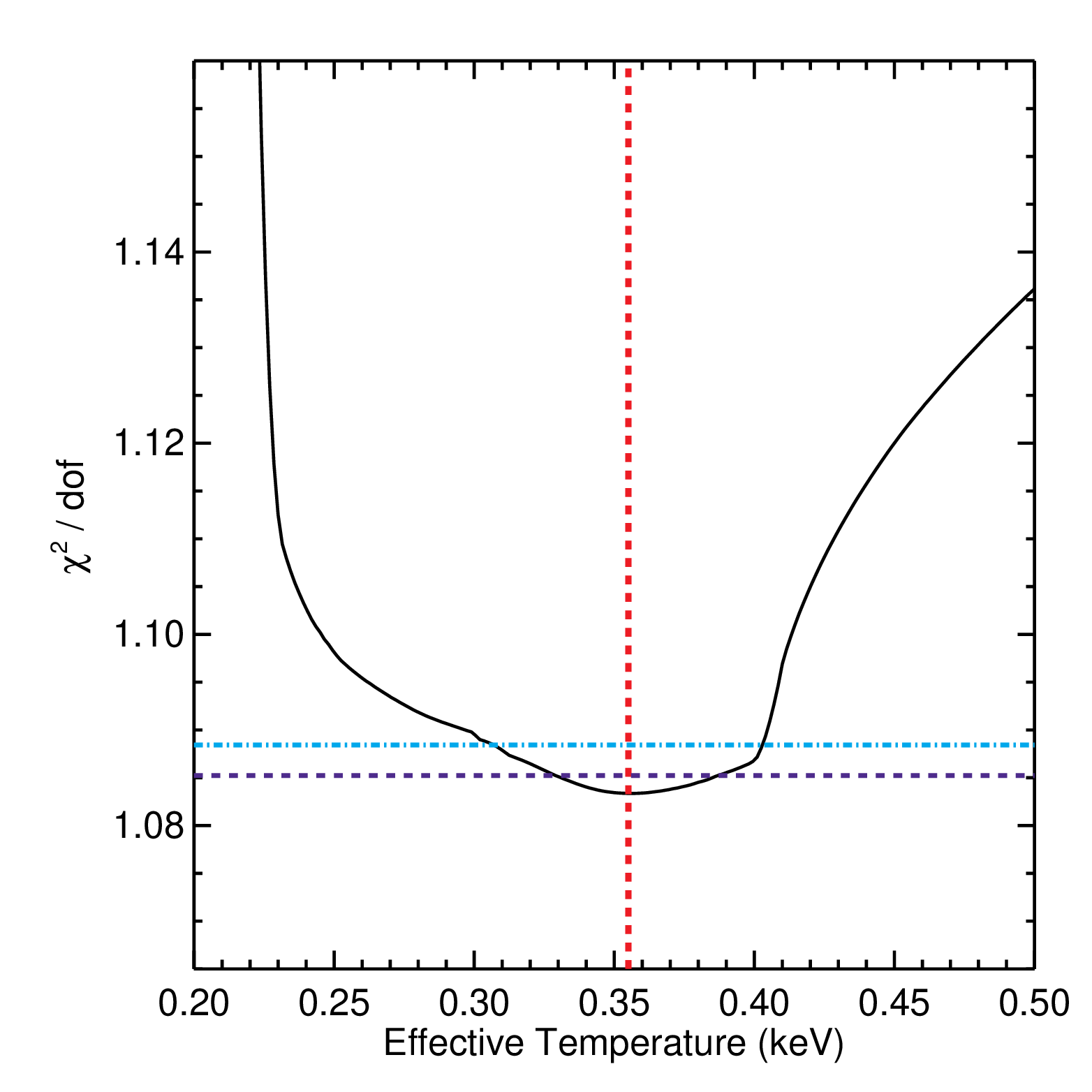}
\includegraphics[scale=0.3]{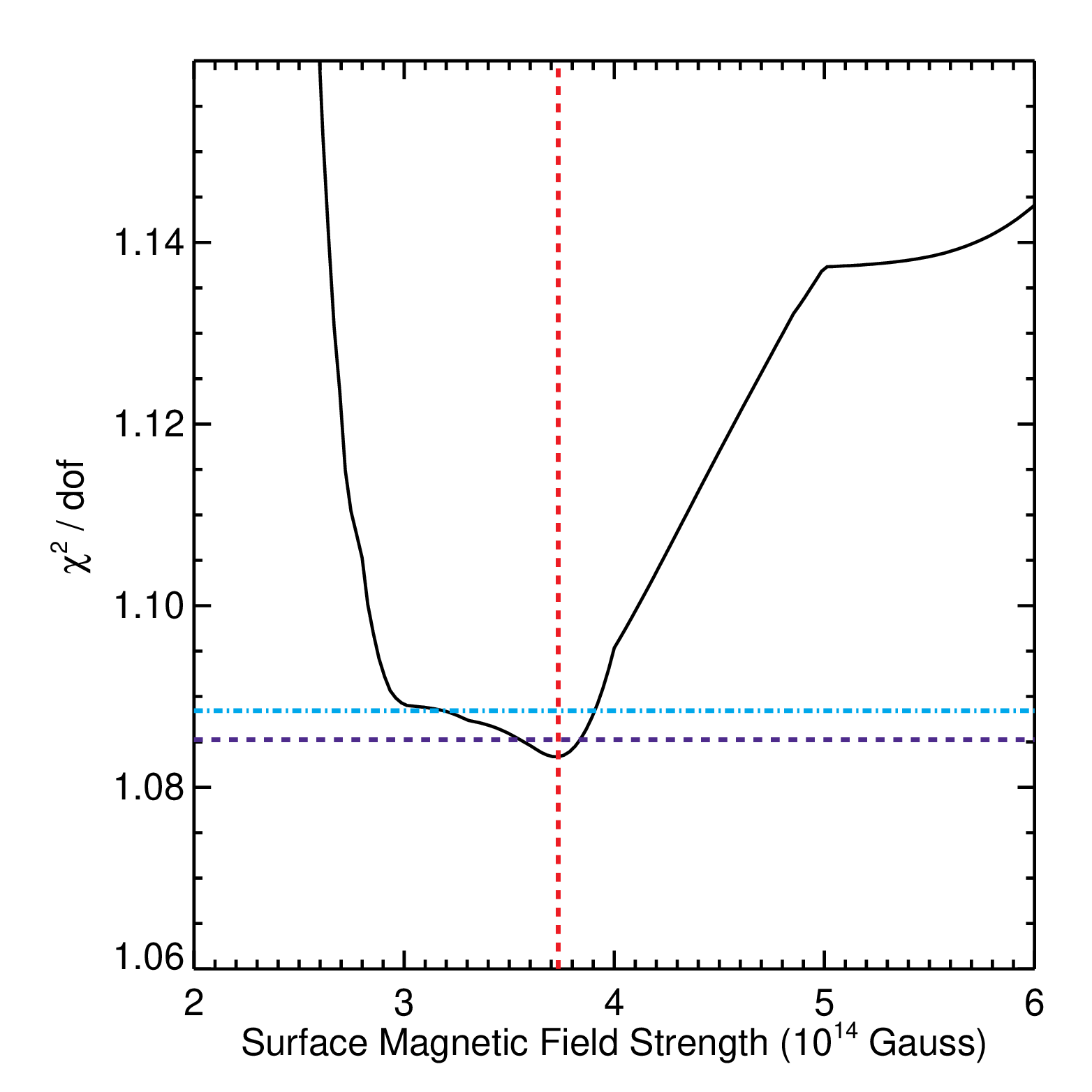}
\caption{Variations  of  $\chi^{2}$/dof   as  a  function  of  surface
  effective  temperature  (left  panel)  and magnetic  field  strength
  (right panel) as  inferred from the X-ray spectral  fit.  Dashed and
  dashed-dotted lines indicate the 68\% and 90\% confidence levels for
  each parameter.}
\label{contours}
\end{figure*}

Given  the fact  that  the surface  temperature  remains constant,  we
repeated  our  simultaneous  fit  by  linking  the  surface  effective
temperature  among  all observations  as  well,  in  order to  further
constrain it and allowed the normalization, hence the emitting area of
the STEMS model, to vary  between observations. This way we obtained a
$\chi^{2}$/dof  of 1.083/535.   Figure \ref{best_fit}  shows  the data
together with the best fitting  model and fit residuals.  As expected,
the best  fit values  are very  similar to what  we found  earlier: We
obtain   that  the  magnetic   field  strength   at  the   surface  is
B$=(3.73^{+0.08}_{-0.16})\times10^{14}$~G,   the   optical  depth   to
resonant   scattering   at   the   neutron   star   magnetosphere   is
$\tau=5.47^{+0.75}_{-0.49}$   and   the   average  velocity   of   the
magnetospheric  particles  is  $\beta=0.52\pm0.03$  and  an  effective
temperature of  0.355$_{-0.024}^{+0.031}$ keV.  We note  that the best
fit value of the surface  temperature does not significantly depend on
the fixed  hydrogen column  density. Even when  we allow  the hydrogen
column density  to be a  free parameter in  the fit to the  spectra of
SGR~0526$-$66,  we obtain a  value of  $N_{\rm H}  = (0.144~\pm~0.012)
\times  10^{22}$~cm$^{-2}$,  which is  consistent  with  the value  we
obtained  from analyzing  the  spectrum of  the surrounding  supernova
remnant. The  unabsorbed flux and the emitting  radius values obtained
from   this  fit  is   given  in   Table  \ref{kt_flux}.    In  Figure
\ref{contours},  we present $\chi^{2}$/dof  contours over  the surface
effective temperature and magnetic field strength.

\section{Timing Analysis}

We performed  timing analysis using all available  CXO observations to
determine the spin period evolution  of \sgr and to uncover variations
in the  pulsed fraction.  The  time resolution of CXO  observations in
various sub-array  modes is approximately 0.4  or 0.8 s,  which is not
ideal but  sufficient for timing  purposes.  Using the  same selection
regions described  in Section  \ref{spec}, we extracted  source events
for each observation and applied barycentric correction using the {\it
  axbary} tool.

We  also included the  deep XMM-Newton  observation performed  in 2007
(ObsID  0505310101) in  our timing  investigation.  We  calibrated the
EPIC-pn  data using  SAS  v.10.0.0  and the  calibration  files as  of
October 2010. We  extracted source events from a  circular region with
10$\arcsec$ radius. Note that  the point spread function of XMM-Newton
is  not accurate  enough to  completely  resolve the  pulsar from  the
supernova remnant;  therefore, some  fraction of unpulsed  emission is
expected to originate from the remnant (see Tiengo et al. 2009 for the
details of contribution from the  supernova remnant). We used the {\it
  barycen} tool of SAS to convert  each event arrival times to that of
the Solar system barycenter.

\begin{figure*}
\centering
\includegraphics[scale=0.35]{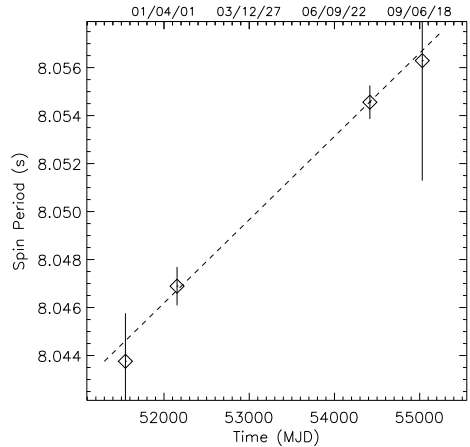}
\caption{Spin period history of SGR~0526$-$66.}
\label{period}
\end{figure*}

To  search for  the pulsed  signal from  SGR~0526$-$66, we  employed a
Z$^{2}_{m}$  technique (Buccheri  et  al.  1983)  with  the number  of
harmonics set  to m  = 2. We  performed the  search in a  period range
between  8.0  and  8.1 s.   We  detect  the  pulsed signal  with  high
significance  in the  first three  CXO  data sets  as well  as in  the
XMM-Newton  data, while  the  detections  in the  CXO  data sets  with
observation  IDs  10806, 10807  and  10808  were  marginal.  In  Table
\ref{timing},   we   present   only  the   statistically   significant
measurements of  the spin period  of SGR~0526$-$66, together  with the
chance probability  and the Z$_{2}^{2}$ power of  each measurement. To
determine the rate  of change of the spin period,  we fit the measured
periods with  a first order polynomial.   We obtain a good  fit with a
spin  down  rate of  (4.019~$\pm$~0.494)$\times$10$^{-11}$~s~s$^{-1}$.
We present  the evolution of the spin  period as well as  the best fit
spin down rate of SGR~0526$-$66  in Figure \ref{period}. Note that our
result regarding the spin evolution is based on only four measurements
obtained over nine years. A  more precise measurement of the spin down
rate require more frequent deep observations.

Finally,  we   constructed  the  pulse  profile   and  calculated  the
root-mean-square pulsed fraction for each observation with significant
spin period detection. The rms pulsed fraction (PF) is calculated as,

\begin{equation}
PF = \left ( \frac{1}{\rm N}  ( \sum_{i=1}^{\rm  N} ( {\rm  R}_{\rm i}-{\rm R}_{\rm ave})^{2}
- \Delta {\rm R}_{\rm i}^2 ) \right )^{\frac{1}{2}} / {\rm R}_{\rm ave},
\end{equation}

where N is the number of pulse phase bins (N=16), ${\rm R}_{\rm i}$ is
the source count  rate in each phase bin, $\Delta  {\rm R}_{\rm i}$ is
the associated uncertainty in the  count rate, and ${\rm R}_{\rm ave}$
is the  average count rate  of the pulse  profile. We present  the rms
pulsed fraction values in Table \ref{timing}.  We find that rms pulsed
fraction of SGR~0526$-$66 is very  low and remains constant around 4\%
among CXO observations, while pulsed fraction obtained from XMM-Newton
observation is significantly  lower ($\sim$1.5\%).  Note, however, the
fact that the  rms pulsed fraction value is  normalized by the average
count  rate of the  pulse profile  and any  blended emission  from the
supernova  remnant would  increase  the average  rate  and reduce  the
pulsed fraction.  It is  likely that the  drop of rms  pulsed fraction
only seen in XMM-Newton observation is not intrinsic to the source.

\begin{table*}
\caption{Results of the timing analysis}  
\begin{tabular}{cccccc}
\hline
Obs ID & Obs Date & P$_{spin}^{a}$ & Maximum Z$_{2}^2$ & Probability$^{b}$ & Pulsed Fraction$^{a}$  \\
       & (MJD)    & (s)            & Power   &   & \\
\hline
747 & 52415.400   &    8.044(2) &   19.796 &
5.5$\times$10$^{-4}$ & 0.038(10) \\
1957 & 52030.200   &    8.0469(8) &  24.518 &
6.3$\times$10$^{-5}$ & 0.042(9) \\
XMM & 55044.000   &    8.0546(7) &  19.027 &
7.8$\times$10$^{-4}$ & 0.015(4) \\
10123 & 55090.500   &    8.056(5) & 15.130 & 
4.5$\times$10$^{-3}$ & 0.044(14) \\
\hline
\end{tabular}
\footnotesize{ 
\begin{flushleft} 
    $^{a}$68\% confidence limits of the measurements are in the
    last digit and given in parentheses. \\
    $^{b}$Probability of obtaining the peak power in the Z$_{2}^2$ power spectrum if there was no periodic signal present.
\end{flushleft}
}
\label{timing}
\end{table*}

\section{Discussion and Conclusions}

We   performed  a   systematic  analysis   of  the   archival  Chandra
observations  of  SGR~0526$-$66.  Fitting  the  X-ray  spectra with  a
strongly magnetized atmosphere model allowed us determine the strength
of  the  surface magnetic  field  as  $B=3.73\times10^{14}$~G and  the
effective  temperature as  0.355~keV.   We also  obtained the  optical
depth to  resonant scattering  in the magnetosphere,  $\tau$=5.47, and
the  average velocity  of  the magnetospheric  particles, 0.52c.   All
these parameters remain constant over  the course of nine years within
uncertainties.

What  has been observed  to change  over this  interval is  the source
flux: it decreased  by about 20\% from 2000 to 2009  (see also Park et
al.   2012). Such  a  flux  decay can  result  from a  $\approx$~5-6\%
decrease in the surface  temperature.  This variation is comparable to
the  uncertainties  in  individual  temperature  measurements  at  the
1-$\sigma$  level.  Therefore,  we can  not unambiguously  rule  out a
variation in  the temperature.  Nevertheless,  no systematic variation
in the  best fit values of  the surface temperature  has been observed
over  the  nine years.   Therefore,  it  is  likely that  the  surface
temperature of \sgr remained constant throughout this period.

The decline in the observed flux  can also be achieved with a decrease
in  the radius  of  the emitting  region  by about  10$\%$. We  obtain
through  our  spectral fits  that  the radius  was  about  13.3 km  in
2000/2001, and  about 11.7  km in 2009,  assuming a distance  48.1 kpc
(Macri et al.  2006).  It is possible that  the thermally emitting hot
spot on the surface of the  neutron star might still carry the effects
of burst induced heating in the  late 70s and early 80s and dissipates
over time without going through a major change in its temperature.  It
is important to  note that the large radius values  we deduce from the
spectral fits are in agreement  with the very low rms pulsed fractions
we inferred from the timing analysis.

We determined  the spin period  history of SGR~0526$-$66, which  is in
accord  with the  results of  Tiengo et  al.\ (2009)  and  derived the
period derivative as 4.02$\times$10$^{-11}$~s~s$^{-1}$.  Assuming that
the spindown of the neutron star  is due to a magnetic dipole braking,
the inferred  dipole magnetic  field strength of  the neutron  star is
calculated as $\approx$~5.7$\times$ 10$^{14}$ G, which agrees with the
surface magnetic field strength as  found from the X-ray spectra using
the STEMS  model.  Making the  same spindown assumption,  we calculate
the characteristic age  of the pulsar as $\tau_{\rm  sd} = P/2 \dot{P}
\approx 3200$~yr. This is comparable  to, but somewhat lower than, the
estimated Sedov age  of the supernova remnant, which  is 4800~yr (Park
et  al.\  2012).  Because  SGRs  and  AXPs  exhibit episodic  variable
spindown behavior (see, e.g.,  Woods et al.\ 2002), the characteristic
age should only be  taken as a rough indicator of the  true age of SGR
0526$-$66. If \sgr is indeed associated with N49, it requires a rather
high radial velocity for the  neutron star compared to some other AXPs
and SGRs (Gaensler et al. 2001; Kaplan et al.  2009).

Long term observations of magnetars  may play an important role in our
understanding  of the  magnetic field  decay  and its  effects on  the
neutron  star crust and  long term  cooling. However,  observations of
magnetars over  the last couple  of decades showed that  these neutron
stars are rarely in quiescent state: They show giant flares (Hurley et
al.  1999;  2005), outbursts (Ng et  al.  2011), glitches  (Dib et al.
2007),  which   likely  affect  the  long-term   cooling  prevent  any
observational  constrains on  the theoretical  calculations.  Although
located at a  large distance and residing within  a SNR, SGR~0526$-$66
is an exception in this regard. Since the giant flare observed in 1979
and the bursting activity that lasted till 1983, no burst activity has
been reported from this source,  enabling us to assume that the source
has been  steadily cooling in  quiescence.  This is also  confirmed by
our  spectral  results presented  in  Section  \ref{spec}.  One  other
advantage of  this source is  its possible association  with the
supernova  remnant N49, providing  a reasonable  age estimate  for the
neutron star.

\begin{figure}
\centering
\includegraphics[scale=0.3]{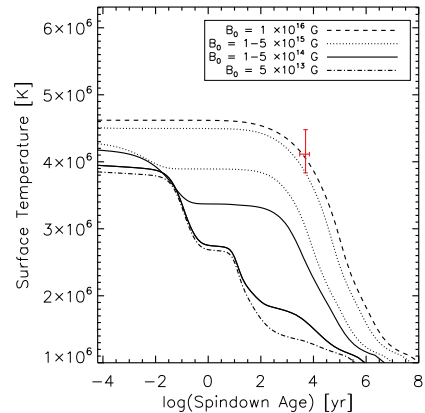}
\caption{Surface temperature  of \sgr  inferred using the  STEMS model
  together with a plausible age  range (red cross sign) overplotted to
  surface temperature  at the pole as  a function of  spindown age for
  magnetars  with  different   initial  magnetic  field  strengths  as
  calculated by Aguilera et al.  (2008a, b).}
\label{cooling} 
\end{figure}

Aguilera  et  al.   (2008a)  calculated cooling  curves  for  strongly
magnetized neutron stars by  taking into account the Ohmic dissipation
($\tau_{\rm Ohm}$) and Hall drift ($\tau_{\rm Hall}$) processes, where
the latter is employed for a  rapid decay during early stages (see eq.
17 in  Aguilera et al.  2008a).   We can compare our  results with the
magnetic neutron  star cooling curves  calculated by Aguilera  et al.\
(2008a, b) by using a conservative age range for SGR 0526-66, together
with  the surface  temperature  that we  determined  from six  Chandra
observations covering nearly a 10  year time interval. We take the age
of the neutron star to be $3000-7000$~yr, which spans the spindown age
of SGR 0526-66 and the supernova  remnant age reported by Park et al.\
(2003, 2012).  In Figure  \ref{cooling}, we present the cooling curves
for  magnetars with  various initial  magnetic field  strengths, along
with the current surface  temperature of SGR~0526$-$66. Our comparison
shows  the  cooling trend  of  SGR~0526$-$66  matches the  theoretical
cooling  curves corresponding to  two sets  of parameters,  an initial
magnetic field  strength of 10$^{16}$ G, $\tau_{\rm  Ohm}$ of 10$^{6}$
yr and $\tau_{\rm Ohm} / \tau_{\rm Hall}$ = 5$\times$10$^2$ yr, or the
curve  that  corresponds to  an  initial  magnetic  field strength  of
1-5$\times$10$^{15}$~G, $\tau_{\rm Ohm}$ of 10$^{6}$ yr and $\tau_{\rm
  Ohm} / \tau_{\rm Hall}$ = 5$\times$10$^3$ yr.

On either  cooling track, \sgr  is expected to  enter a phase  of more
rapid cooling  in the near  future, based on  its age and  its current
temperature, which we reported  here. If its temperature indeed begins
to decay rapidly, observations of the source over the next decade with
planned X-ray  telescopes, such as Advanced Telescope  for High Energy
Astrophysics,  may reveal  a  more significant  drop  in the  spectral
temperature,  accompanied by  a further  flux  decay, as  long as  the
source remains in its deep quiescence.

\section*{Acknowledgments}
We thank Deborah Aguilera and  Jose Pons for sharing their theoretical
calculations on  the cooling  of magnetars. We  thank the  referee for
insightful suggestions  that improved  the clarity of  the manuscript.
TG acknowledges support from the Scientific and Technological Research
Council   of  Turkey  (T\"UB\.ITAK   B\.IDEB)  through   a  fellowship
programme.

\end{document}